\documentclass[a4paper]{jpconf}
\usepackage{graphicx}
\begin{document}
\title{The propagator for the step potential using the path decomposition expansion}
\author{James M Yearsley}
\address{Blackett Laboratory, Imperial College, London SW7 2BZ, UK }
\ead{james.yearsley@imperial.ac.uk}

\begin{abstract}
We present a direct path integral derivation of the propagator in the presence of a step potential. The derivation makes use of the Path Decomposition Expansion (PDX), and also of the definition of the propagator as a limit of lattice paths. 
\end{abstract}

\section{Introduction}
The propagator for the step potential, $V(x)=V_{0}\Theta(-x)$, is a simple model for scattering, and a useful ingredient in arrival time problems \cite{halliwell}. Expressions for this propagator are known \cite{CO}, but existing derivations are complex, and give little insight into the structure. Here we show how this propagator may be obtained in a natural way via the Path Decomposition Expansion (PDX), in which the propagator is factored across a surface of constant time \cite{pdx} and the Brownian motion definition of the path integral. \footnote{This contribution is based on \cite{me}.}

This propagator may be viewed as a conditional probability density for a random walk on the real line. We then make this integral over paths into a concrete object by defining it as the continuum limit of a discrete sum of lattice paths, $u(0,T|0,0)$.
 We consider a rectangular lattice with spacing in the time direction of $\epsilon$, and spacing in the $x$ direction of $\eta$, and consider propagation for a time $T=2\epsilon n$, so we have $2n$ steps in our paths. 
 The Euclidean propagator $\overline g$ is then defined as the continuum limit of $u/2\eta$ where we take $\epsilon,\eta \rightarrow 0$, $n\rightarrow \infty$, keeping $\epsilon/\eta^{2} = m$ and $T=2\epsilon n$ fixed. That is,
\begin{equation}
\overline g(0,T|0,0):=\lim_{\eta, \epsilon \to 0} (2\eta)^{-1}u(0,T|0,0)\nonumber.
\end{equation}

\section{The Path Decomposition Expansion (PDX)}
The PDX is a very useful tool for evaluating discontinuous potentials, such as the step potential. A typical path from $x_{0}<0$ to $x_{1}$ may cross $x=0$ many times, but the set of paths may be partitioned 
according to their {\em first} and {\em last} crossing time. We therefore split every path into three parts: (A) a restricted part that starts at $x_{0}$ and does not cross $x=0$, but that ends on $x=0$ at time $t_{1}$, (B) an unrestricted part from $x=0$ to $x=0$ that may cross $x=0$ many times and, (C) a further restricted part from $x=0$ to $x_{1}$ that does not re-cross $x=0$, see Fig. 1.
\begin{figure}
\begin{center}
\includegraphics[]{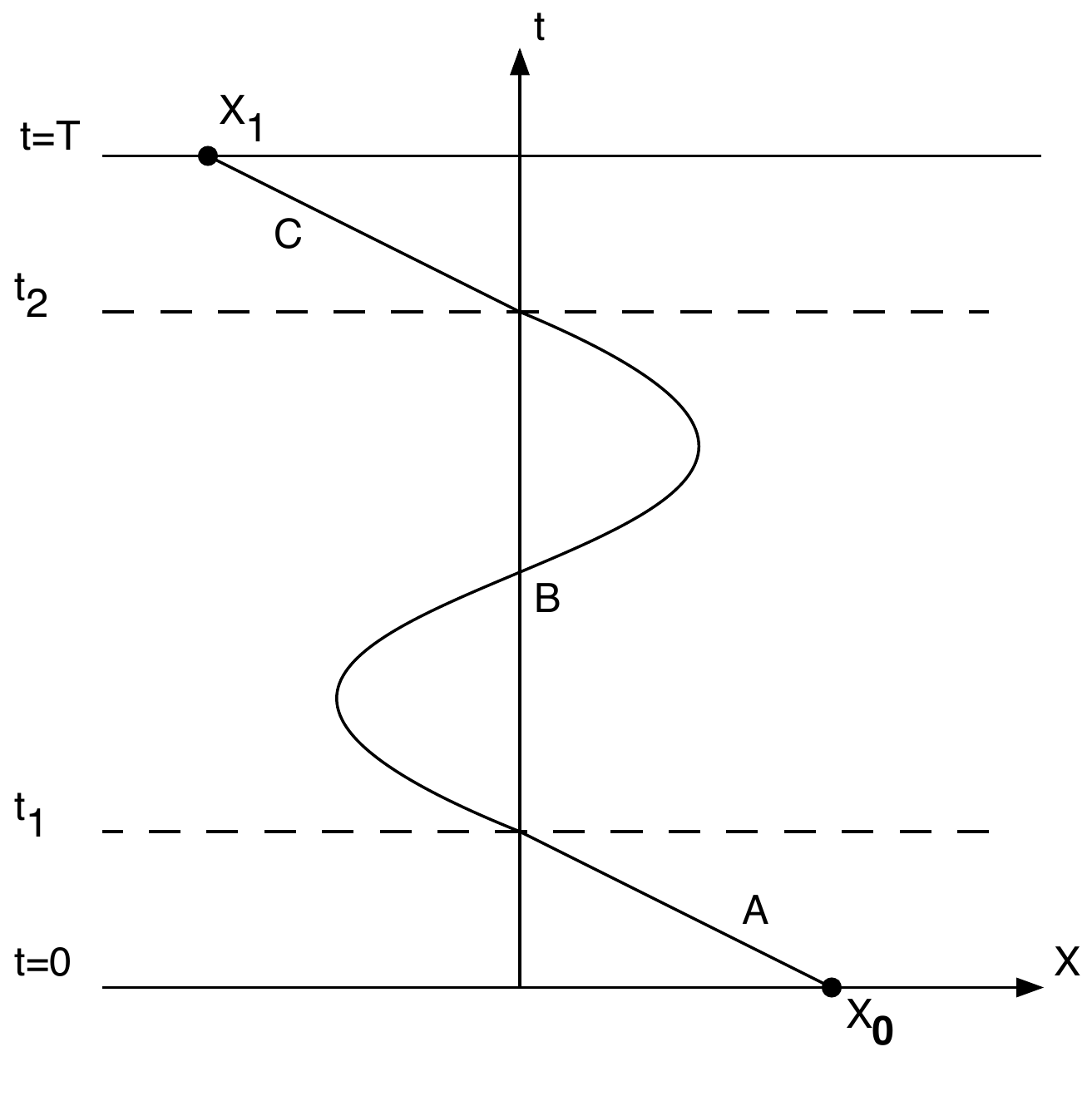}
\caption{A typical path from $x_{0}$ to $x_{1}$.}
\end{center}
\end{figure}
The propagator can therefore be written as \cite{pdx};
\begin{eqnarray}
 g(x_{1},T|x_{0},0)& =&   \frac{\hbar^{2}}{4m^{2}}\int_{0}^{T}dt_{2}\int_{0}^{t_{2}}dt_{1}   \;  \frac{\partial g_{r}}{\partial x}(x_{1},t|x,t_{2})|_{x=0}    \; g(0,t_{2};0,t_{1}) \nonumber \\
 & & \times  \; \frac{\partial g_{r}}{\partial x}(x,t_{1};x_{0},0)|_{x=0}\label{path}.\nonumber
\end{eqnarray}
Where $ t_{1}$ and $t_{2}$ are the first and last crossing times and the $g_{r}$ are the restricted propagators for the regions $x<0$ and $x>0$, where the potential is constant. They are therefore easily computed by the method of images.
Note that when $x_{1}$ and $x_{0}$ are both positive, there is an additional contribution from paths which never cross $x=0$ \cite{pdx}.
The calculation of the propagator therefore reduces to the much simpler task of computing $g(0,T|0,0)$, the propagator along the {\em edge} of a step potential.

\section{Lattice Paths}
We first switch to working with the Euclidean propagator $\overline g$ by means of a Wick rotation and we specialise immediately to the case of $x_{0}=x_{1}=0$.  That is, we wish to calculate
\begin{equation}
\overline g(0,T | 0,0)=\int_{x(0)=0}^{x(T)=0}\! \mathcal{D}x \;e^{-S_{E}/\hbar}\label{prop}
\end{equation} 
where $S_{E}$ is the Euclidean action given by
 \begin{equation}
S_{E}=\int_{0}^{T}dt\left(\frac{m\dot x^{2}}{2} + V_{0}\Theta(x)\right)\nonumber.
\end{equation}

On the lattice the conditional probability $u(0,T|0,0)$ is given by a sum of paths, each weighted by $e^{-V_{0}\tau}$, where $\tau$ is the length of time spent in $x<0$.
This may be written as
\begin{equation}
u(0,T|0,0) = \frac{1}{2^{2n}}\sum_{k=0}^{n} \;n_{k}\; e^{-2k\epsilon V_{0}}\nonumber,
\end{equation}
where $n_{k}$ is the number of paths spending a time $2k\epsilon<T$ in the region $x<0$. Expresions for these $n_{k}$ are known and are in fact independent of $k$ and are equal to the Catalan numbers \cite{me, combinatorics}.
\begin{equation}
C_{n} = \frac{1}{n+1} {2n \choose n}\nonumber
\end{equation}
where $2n$ is the total number of time steps.  This is the central observation that allows this propagator to be computed in closed form.

The sum can therefore be performed and, using the very useful asymptotic form for the $C_{n}$

\begin{equation}
C_{n}\approx \frac{2^{2n}}{\sqrt{\pi}n^{3/2}}\nonumber,
\end{equation}

we find
\begin{displaymath}
u(0,T|0,0) \approx  \frac{1}{\sqrt{\pi}2\epsilon n^{3/2}}\left(\frac{1-e^{-2\epsilon(n+1)V_{0}}}{V_{0}}\right).
\end{displaymath}
Taking the continuum limit and Wick rotating back to real time then yields the final result,
\begin{equation}
g(0,T|0,0) = -i \left(\frac{m}{2 \pi i}\right)^{1/2}\frac{(1-e^{-iV_{0}T})}{V_{0}T^{3/2}},\nonumber
\end{equation}
which is the expression for the propagator along the edge of a step potential. The full propagator from $x_{0}$ to $x_{1}$ may now be obtained using the PDX \cite{me}.

\section{Summary}
The PDX, together with some elementary combinatorics, allows for a simple derivation of the propagator for this potential. A similar method allows for the computation of the propagator for a delta function potential, by counting the number of crossings of the origin \cite{me}. It is interesting to note that the method described reduces the computation of a path integral to a problem in combinatorics, and this suggests that other propagators may be computed in this way.

\ack

The author would like to thank the organisers of DICE 2008, especially Hans-Thomas Elze, for the opportunity to to take part in this conference. He is also indebted to Jonathan J Halliwell for his support, and for many useful discussions.

\end{document}